\documentclass[aps,prl,twocolumn,english,balance,superscriptaddress,floats,showpacs,longbibliography,nofootinbib]{revtex4-2}
\usepackage[T1]{fontenc}
\usepackage[latin9]{inputenc}
\usepackage{amsmath}
\usepackage{amssymb}
\usepackage{stmaryrd}
\usepackage{graphicx}
\usepackage{esint}
\usepackage{subfigure}
\usepackage{multirow}
\usepackage{wasysym}
\usepackage{xcolor}
\usepackage{mathtools}
\usepackage{gensymb}

\makeatletter

\newcommand{\beq}{\begin{equation}}
\newcommand{\eeq}{\end{equation}}
\newcommand{\bea}{\begin{eqnarray}}
\newcommand{\eea}{\end{eqnarray}}
\newcommand{\bwt}{\begin{widetext}}
\newcommand{\ewt}{\end{widetext}}
\@ifundefined{textcolor}{}
{%
 \definecolor{BLACK}{gray}{0}
 \definecolor{WHITE}{gray}{1}
 \definecolor{RED}{rgb}{1,0,0}
 \definecolor{GREEN}{rgb}{0,1,0}
 \definecolor{BLUE}{rgb}{0,0,1}
 \definecolor{CYAN}{cmyk}{1,0,0,0}
 \definecolor{MAGENTA}{cmyk}{0,1,0,0}
 \definecolor{YELLOW}{cmyk}{0,0,1,0}
}

\newcommand{\bv}{\mathbf{v}}
\newcommand{\bE}{\mathbf{E}}

\newcommand{\bk}{\mathbf{k}}

\newcommand{\bq}{\mathbf{q}}
\newcommand{\bp}{\mathbf{p}}
\newcommand{\br}{\mathbf{r}}

\usepackage{babel}
\makeatother
\usepackage{babel}
\begin{document}

\title{Weak localization and antilocalization corrections to nonlinear transport: a semiclassical  Boltzmann treatment}

\author{Dmitry V. Chichinadze}
\email{cdmitry@wustl.edu}
\affiliation{Department of Physics, Washington University in St. Louis, St. Louis, Missouri 63160, USA}
\affiliation{National High Magnetic Field Laboratory, Tallahassee, Florida, 32304, USA}

\begin{abstract}
    The nonlinear transport regime is manifested in the nonlinear current-voltage characteristic of the system. An example of such a nonlinear regime is a setup in which current is injected into the sample and the measured voltage drop is quadratic in the injected current. Such a quadratic nonlinear regime requires inversion symmetry to be broken. This is the same symmetry condition as one needs to observe weak antilocalization, which can be prominent in two-dimensional systems. Here, we study the effects of weak (anti)localization on second-order nonlinear transport in two-dimensional systems using the semiclassical Boltzmann approach. We solve for quasiparticle distribution function up to the second order in the applied external electric field and calculate linear and nonlinear conductivity tensors for a toy model. We find that localization effects could lead to a sign change of the nonlinear conductivity tensor -- a phenomenon observed in transition metal dichalcogenide and in single-layer graphene devices. 
\end{abstract}

\maketitle

\section{Introduction}

The problem of nonlinear transport in systems lacking inversion symmetry is long-standing \cite{GenkinMednis1968}. A significant fraction of recent progress in this field relied heavily on the use of semiclassical Boltzmann approach \cite{Gao2014PRLNHEmetric,Sodemann_Fu,KoenigDzeroLevchenkoPesin,IsobeSciAdv2020,Wang2021PRL,AgarwalQG2023PRB}. Given the overall advancement of the field, this approach left certain questions unanswered. Specifically, the question of the physical origins of giant nonlinear conductivity in layered graphene-based van der Waals heterostructures remains open. This question was highlighted by an apparent discrepancy (up to several orders in magnitude) between the theoretically expected values of nonlinear conductivity and the experimentally measured components of nonlinear conductivity tensor, extracted both in Hall bar- \cite{He2022} and in disk-shaped \cite{Chichinadze2025GiantNonlinearHall} samples. This enormous discrepancy strongly hints that our understanding of the microscopic origins of nonlinear conductivity in 2D van der Waals heterostructures is incomplete and requires re-evaluation. 

One of the possible reasons for such an incomplete understanding is the limitations in the theoretical treatment of the problem. On the one hand, many theoretical calculations are done using a semiclassical Boltzmann approach in the relaxation time approximation. Such an approach, while capable of providing valuable insights into the phenomenology and giving an intuitive understanding of the underlying physics, is severely incomplete, as it averages out finer details of the system and largely disregards important quantum effects. An alternative to this is a full quantum kinetic equation and/or a proper treatment of impurity scattering \cite{Du2021NatComm}, which is tedious and seems to provide an incremental understanding of the microscopic effects driving the nonlinear current response of the system. On the other hand, even certain quantum effects that can be qualitatively treated within the semiclassical Boltzmann approach are often disregarded. 

In this work, we consider one of such effects, namely weak localization and antilocalization, using a semiclassical Boltzmann approach \cite{AmbegaokarPRB1986,KimPRB2014,KimPRB2018}. While the Boltzmann approach is overly simplistic to properly treat such quantum mechanical effects \cite{Schwab2001,McCannWeakLocalization2006PRL,Khveshchenko2006,GarateGlazmanPRB2012,KimPRB2014,Shen2015PRB,KimPRB2018}, it is important to illustrate the consequences of these effects even on the semiclassical level, as they could play an important role in our understanding of nonlinear transport in 2D systems. To make our point transparent, we limit ourselves strictly to the effects of weak (anti)localization and do not discuss the implications of corrections arising from broken time-reversal symmetry. As such, we leave anomalous velocity out of consideration.  We also leave out of consideration the question of changing local temperature, brought up in, e.g., Refs. \cite{AmbegaokarPRB1986,Leadbeater2000,BelitzPRB2004} and potential quantum geometric effects.

From an experimental perspective, localization and antilocalization effects in semiconductor heterostructures are known to be present, and their interplay is dependent on the strength of spin-orbit coupling, which can be thought of as a measure of inversion symmetry breaking. For instance, top gate voltage control was shown to move the system from weak localization to a weak antilocalization regime in semiconductor heterostructures with spin-orbit coupling \cite{Gossard2003PRL,Nitta2006Crossover}. The same effect was also observed in single-layer graphene samples \cite{GrapheneLocalizationExperiment2008PRL,Gorbachev2009transition,Gopinadhan2013}. Displacement-field-influenced SOC was shown to be present in first principles calculations of graphene electronic band structure \cite{Gmitra2009}, as well as in experiments on TMD-proximitized bilayer graphene \cite{AmannPRB2022} and single-layer graphene flakes \cite{Gopinadhan2013}.
From the fundamental perspective, the broken inversion symmetry is the necessary condition to observe both nonlinear transport \cite{SuarezRodriguez2025} and weak antilocalization \cite{HikamiLarkinNagaoka}; hence, without knowing the actual strength of the quantum interference effects, one cannot immediately rule them out of consideration. 

The key findings of this work are that the effects of weak (anti)localization can lead to (i) a rise in magnitude and (ii) to a sign change of nonlinear conductivity in the absence of electronic transitions in the system. The nonlinear conductivity sign change, as a function of electron density, can be seen in the transport measurements data on transition metal dichalgocenides \cite{Ma2019} and on single-layer graphene heterostructures \cite{He2022}. The significant rise in magnitude might be outside the applicability range of our theory and requires further microscopic treatment. 


The structure of the paper is as follows. In Sec. \ref{sec:kinetics} we introduce the Boltzmann equation formalism, taking into account weak (anti)localization, and derive expressions for linear and nonlinear conductivities in the relaxation time approximation. In Sec. \ref{sec:min_model} we consider a minimal model as an example and discuss how nonlinear conductivity is affected by weak (anti)localization. We discuss experimental relevance and conclude in Sec. \ref{sec:experiment_and_conclusions}.

\section{Kinetic equation and conductivity tensors}
\label{sec:kinetics}

To study the effects of weak (anti)localization on nonlinear conductivity, we employ the Boltzmann equation in relaxation-time approximation but with time-nonlocal collision integral, following \cite{AmbegaokarPRB1986,KimPRB2014,KimPRB2018}:
\begin{widetext}
\begin{equation}
    \begin{gathered}
        \left( \frac{\partial}{\partial t} + \dot{\br} \cdot \nabla_{\br} + \dot{\bp} \cdot \nabla_{\bp} \right) f(\bp, \br, t) = -\frac{f(\bp, \br, t) - f_{\text{eq}}(\bp)}{\tau} + \int_{- \infty}^{t} dt' \alpha \left( t - t' \right) \left[  f(-\bp, \br, t') - f_{\text{eq}}(\bp) \right],
    \end{gathered}
    \label{Full_Boltzmann}
\end{equation}
\end{widetext}
where the units are set such that $\hbar=1$, $f(\bp, \br, t)$ -- is the non-equilibrium quasiparticle distribution function, $f_{\text{eq}}(\bp)$ -- is the equilibrium distribution function given by the Fermi-Dirac distribution at zero temperature, $\tau$ -- is the relaxation time, and $\alpha \left( t - t' \right)$ -- is the time-nonlocal contribution that can be viewed as a diffusion kernel \cite{AmbegaokarPRB1986,KimPRB2014} and in Fourier-space is given by
\begin{equation}
    \alpha (\omega) = \pm \frac{\mathcal{C}}{\pi \nu_{F} \tau} \int_{1/l_{\phi}}^{1/l} \frac{d^2 \bq}{(2 \pi)^2} \frac{1}{D \bq^2 - i \omega }.
    \label{alpha_eq_main}
\end{equation}
Here $\nu_F$ -- is the DOS at the Fermi level, $D$ -- is the diffusion constant, $l$ -- is the mean-free path, $l_{\phi}$ -- is the phase-coherence length, and $\mathcal{C}$ -- is a numerical prefactor that accounts for spin and/or isospin structure of the system. For the standard spinful case, $\mathcal{C}=1$ for weak localization and $\mathcal{C}=\frac{1}{2}$ for antilocalization \cite{HikamiLarkinNagaoka}. However, the value of $\mathcal{C}$ for antilocalization case can differ from the expected $\frac{1}{2}$ in special cases \cite{Ghaemi2010PRL,WLWALCompetition2011PRL}.  The $+$ sign in Eq. \eqref{alpha_eq_main} corresponds to weak localization, while $-$ corresponds to weak antilocalization \cite{KimPRB2014}. As was shown in \cite{AmbegaokarPRB1986,KimPRB2014,KimPRB2018}, Eq. \eqref{Full_Boltzmann} recovers known results for weak (anti)localization in the linear response regime. In what follows, we will restrict ourselves to the spatially homogeneous case and assume $f(\bp, \br, t) = f(\bp, t)$.

We consider a system of electrons with broken inversion symmetry. In such a case, weak antilocalization in the system is allowed \cite{HikamiLarkinNagaoka} and can lead to \emph{increasing} conductivity due to increasing rate of forward scattering, in contrast to the back scattering reducing conductivity for weak localization. 
In contrast to \cite{KimPRB2014}, we will not take advantage of time-reversal symmetry as we want to consider the case of a minimal model, which mimics a single valley of graphene, as an example. Performing the standard Fourier transformation $f(\bp, t) = \int_{-\infty}^{\infty} d \omega f(\bp, \omega) e^{- i \omega t} $ and using $\dot{\bp} = e \bE$ allows to cast Eq. \eqref{Full_Boltzmann} into 
\begin{equation}
    \begin{gathered}
        \left( -i \omega + e \bE \cdot \nabla_{\bp} \right) f(\bp, \omega) = \\ = -\frac{f(\bp, \omega) - f_{\text{eq}}(\bp)}{\tau} + \alpha \left( \omega \right) \left[  f(-\bp, \omega) - f_{\text{eq}}(\bp) \right],
    \end{gathered}
\end{equation}
which now will be solved perturbatively in powers of the external electric field $\bE$. Specifically, we expand the non-equilibrium distribution function in a series with $f(\bp, t) = f_0(\bp, t) + f_1(\bp, t) + f_2(\bp, t)$, where $f_{n} \sim (\bE)^n$.
One can then check by a direct substitution that up to the second order in $\bE$ 
\begin{eqnarray}
    f_0 (\bp, \omega) &&= \frac{1}{2} \left[ \frac{\frac{1}{\tau} - \alpha(\omega) }{- i \omega + \frac{1}{\tau} - \alpha(\omega) } \left( f_{\text{eq}}(\bp) +  f_{\text{eq}}(-\bp) \right) + \right. \nonumber  \\ && \left. + \frac{ \frac{1}{\tau} - \alpha(\omega) }{- i \omega + \frac{1}{\tau} + \alpha(\omega) } \left( f_{\text{eq}}(\bp) - f_{\text{eq}}(-\bp) \right) \right], \\
    f_1 (\bp, \omega) &&= - \frac{e \bE \cdot \nabla_{\bp}}{2} \left[ \frac{f_0 (\bp, \omega) - f_0 (-\bp, \omega)}{- i \omega + \frac{1}{\tau} - \alpha(\omega)  } + \right. \nonumber  \\ && \left. + \frac{f_0 (\bp, \omega) + f_0 (-\bp, \omega)}{- i \omega + \frac{1}{\tau} + \alpha(\omega)  } \right], \\
    f_2 (\bp, \omega) &&= - \frac{e \bE \cdot \nabla_{\bp}}{2} \left[ \frac{f_1 (\bp, \omega) - f_1 (-\bp, \omega)}{- i \omega  + \frac{1}{\tau} - \alpha(\omega)  } + \right. \nonumber  \\ && \left. + \frac{f_1 (\bp, \omega) + f_1 (-\bp, \omega)}{- i \omega + \frac{1}{\tau} + \alpha(\omega)  } \right].
\end{eqnarray}
Assuming $f(\bp, \omega) = f(-\bp, \omega)$ one recovers the result for $f_1 (\bp, \omega)$ reported earlier in  \cite{AmbegaokarPRB1986,KimPRB2018}. In the DC limit, which we consider in this work, $\omega \to 0^{+}$ and distribution functions acquire a simple form
\begin{eqnarray}
    &&f_0 (\bp) = \frac{\frac{1}{\tau} f_{\text{eq}}(\bp) + \alpha(0) f_{\text{eq}}(-\bp)  }{\frac{1}{\tau} + \alpha(0) }, \\
    &&f_1 (\bp) = -\frac{e \bE \cdot \nabla_{\bp} }{\frac{1}{\tau} + \alpha(0)} f_{\text{eq}} (\bp),  \\
    &&f_2 (\bp) = - \frac{e^2 E_{\beta} E_{\gamma} \frac{\partial}{\partial p_{\beta}} \frac{\partial}{\partial p_{\gamma}} \left[ \frac{1}{\tau} f_{\text{eq}} (\bp) + \alpha(0) f_{\text{eq}} (-\bp) \right] }{\left( -\frac{1}{\tau} + \alpha(0)  \right) \left( \frac{1}{\tau} + \alpha(0)  \right)^2  },
\end{eqnarray}
where $\lambda, \beta, \gamma = x,y$ -- are spatial indices and $\alpha(0)$ -- is a constant which, together with $\tau$, we treat as a control parameter of our theory. In the limit of $\alpha(0) \to 0$, one retrieves the well-known results $f_1 (\bp) = -e \tau E_{\beta} \frac{\partial}{\partial p_{\beta}} f_{\text{eq}} (\bp)$ and $f_2 (\bp) = e^2 \tau^2 E_{\beta} E_{\gamma} \frac{\partial}{\partial p_{\beta}} \frac{\partial}{\partial p_{\gamma}}   f_{\text{eq}} (\bp)$ \cite{GenkinMednis1968,Wang2021PRL,Chichinadze2025GiantNonlinearHall}. 

Using the standard definition of DC current density for one flavor of spinless fermions
$$
j_{\lambda} = e \int \frac{d^2 \bp}{(2 \pi)^2} v_{\lambda} f(\bp)
$$
we calculate linear and nonlinear conductivity tensors 
\begin{equation}
    \begin{gathered}
        \sigma_{\lambda \beta} = - \frac{e^2}{ \frac{1}{\tau} + \alpha(0)} I^{(1)}_{\lambda \beta}, \\
        \tilde{\sigma}_{\lambda \beta \gamma} = \frac{e^3}{\left( \frac{1}{\tau} - \alpha(0)  \right) \left( \frac{1}{\tau} + \alpha(0) \right)^2  } \left[ \frac{1}{\tau} I^{(2,1)}_{\lambda \beta \gamma}  + \alpha(0) I^{(2,2)}_{\lambda \beta \gamma} \right].
    \end{gathered}
    \label{conductivities}
\end{equation}
Here $\sigma_{\lambda \beta}$ and $\tilde{\sigma}_{\lambda \beta \gamma}$ are linear and nonlinear conductivity tensors respectively, and we introduced dispersion-specific integrals 
\begin{eqnarray}
        I^{(1)}_{\lambda \beta} &=& \int \frac{d^2 \bp}{(2 \pi)^2} \frac{\partial \varepsilon_{\bp}}{\partial p_{\lambda}} \frac{\partial \varepsilon_{\bp}}{\partial p_{\beta}} \frac{\partial f_{\text{eq}}(\varepsilon)}{\partial \varepsilon}, \nonumber \\
    I^{(2,1)}_{\lambda \beta \gamma} &=& \int \frac{d^2 \bp}{(2 \pi)^2} \frac{\partial \varepsilon_{\bp}}{\partial p_{\lambda}} \frac{\partial^2 f_{\text{eq}} (\bp)}{\partial p_{\beta} \partial p_{\gamma}},  \\
    I^{(2,2)}_{\lambda \beta \gamma} &=& \int \frac{d^2 \bp}{(2 \pi)^2} \frac{\partial \varepsilon_{\bp}}{\partial p_{\lambda}} \frac{\partial^2 f_{\text{eq}} (-\bp)}{\partial p_{\beta} \partial p_{\gamma}}, \nonumber
\end{eqnarray}
which explicitly depend on electronic dispersion $\varepsilon_{\bp}$ of charge carriers.
From Eq. \eqref{conductivities} one can immediately notice that $\tilde{\sigma}_{\lambda {\beta} {\gamma}}$ changes sign as a function of $\alpha(0)$ for a fixed $\tau$. This happens because in the relaxation time approximation it is the odd in $\bp$ contribution to $\varepsilon_{\bp}$ that gives rise to any nonlinear conductivity (which is forbidden by symmetry if the rotation symmetry is higher than 3-fold).

\begin{figure*}[t!]
\includegraphics[width=0.99\linewidth]{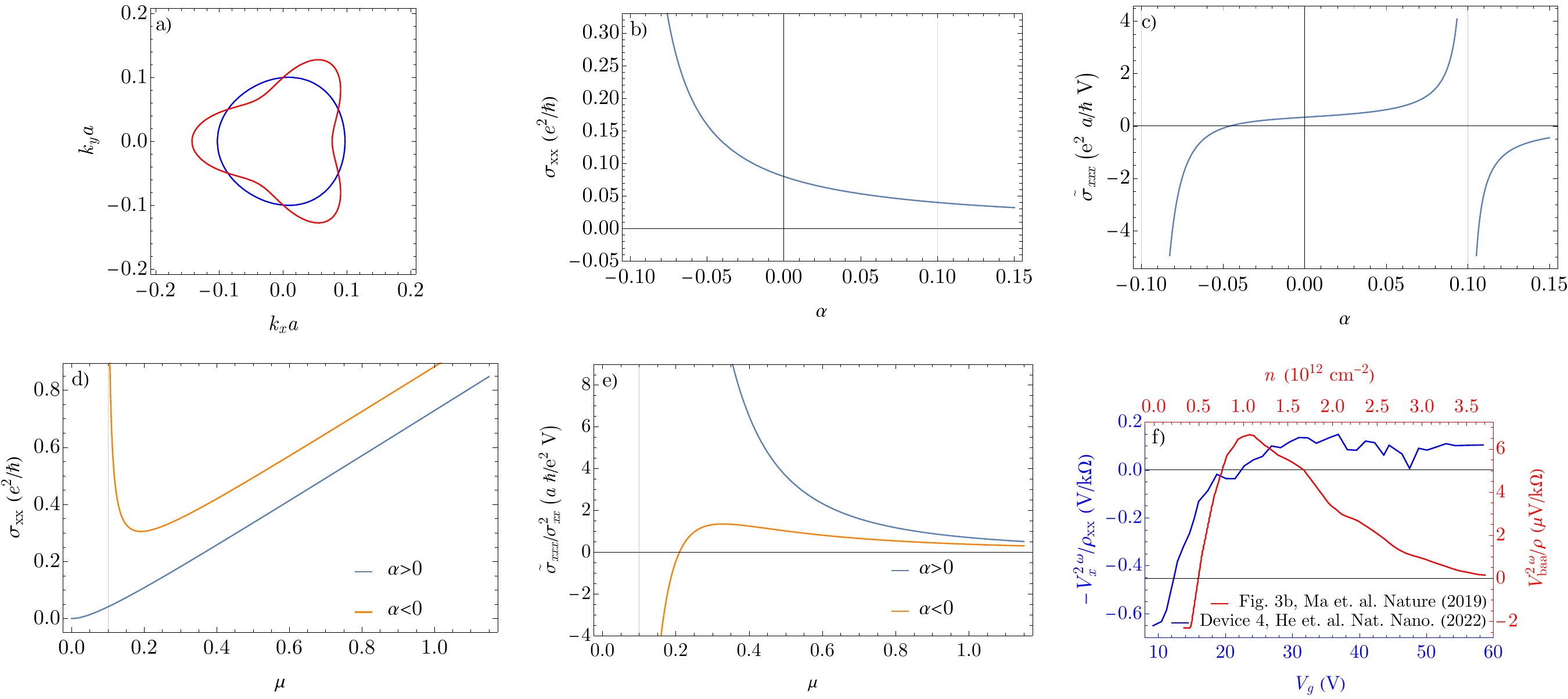}
\centering{}\caption{Results for the minimal model and comparison with experiment: a) Fermi surface contours for the minimal model for $v_D = 1, \mu = 0.1$. For the blue curve $c=0.03$, whereas for the red curve $c=0.3$. The momentum is measured in units of inverse lattice spacing. We set the lattice constant $a=1$. b) Dependence of linear conductivity $\sigma$ Eq. \eqref{lincondres} on $\alpha = \alpha(0)$ calculated for $\tau=10$. Two gray grid lines indicate $\alpha = \pm \tau$. Note that linear conductivity is not well defined for $\alpha < -0.1$ here as it becomes negative. c) Nonlinear conductivity $\tilde{\sigma} (\alpha)$ (see Eq. \eqref{nonlincondres}) for the same value of $\tau$. Note the two types of discontinuities: for $\alpha=-0.1$ both linear and nonlinear conductivities diverge, while for $\alpha = 0.1$ only nonlinear conductivity is divergent. The discontinuity of $\tilde{\sigma} (\alpha)$ for positive $\alpha$ (weak localization case) has a similar shape to the Hall number discontinuity across the Van Hove singularity doping. Importantly, in the weak antilocalization regime $\alpha<0$, nonlinear conductivity changes sign for intermediate values of $\alpha$. In both plots we used $v_D = 1, \mu=0.1, c = 3.3 \times 10^{-2}$. d) Linear conductivity dependence on chemical potential for $\alpha>0$ (localization) and $\alpha<0$ (antilocalization) for $\mathcal{C}=1, D=1$, and $\ln \frac{l_{\phi}}{l} = 0.3$. The unphysical discontinuity in the antilocalization case at small $\mu$ is due to $\alpha \simeq 1/\tau$. For such values of $\mu$, our simplistic description clearly breaks down. e) Plot of ratio $\tilde{\sigma}/\sigma^2$ as a function of chemical potential for the same values of $\mathcal{C}, D, \ln \frac{l_{\phi}}{l}$. Note the sign change for $\mu \sim 0.2$, which is away from the unphysical value regime of divergent linear conductivity. f) Experimental data extracted from Refs. \cite{He2022} and \cite{Ma2019}. For data from \cite{He2022} (shown in blue), nonlinear voltage drop $V^{2 \omega}_{xx}$  is divided by linear resistivity $\rho_{xx}$ as a function of gate voltage $V_g$ which controls electrochemical potential in the system. As we show in Appendix \ref{app:data_an}, $V^{2 \omega}_{xx}/\rho_{xx} \propto \tilde{\sigma}/\sigma^2$. The same data analysis procedure, but for voltage drop $V^{2 \omega}_{baa}$, is applied to the data from Ref. \cite{Ma2019} (shown in red). Note the sign change of the signal and the overall similarity of both curves' shapes to the orange curve in panel e). 
} 
\label{fig1}
\end{figure*}

\section{Theoretical results for a minimal model}
\label{sec:min_model}

To gain more insights on the role of weak (anti)localization on nonlinear conductivity, we calculate linear and nonlinear conductivity dependence on parameter $\alpha=\alpha(0)$ for a minimal model, which describes a spinless Dirac fermion with a trigonal warping term, specified by a parameter $c$:
\begin{equation}
\varepsilon_{\bp} = v_D |\bp| + c |\bp| \cos 3\theta.
\label{disp}
\end{equation}
The presence of the trigonal warping term is crucial as it is the $\cos 3 \theta$ contribution that leads to non-vanishing $\tilde{\sigma}_{\lambda {\beta} {\gamma}}$, as can be verified by a direct calculation, see SI. We show two representative Fermi surfaces of the minimal model in Fig. \ref{fig1}a. 

In this work we consider the small trigonal warping limit, i.e., $\frac{c}{v_D} \ll 1$. In such case, the anisotropy of the scattering rate stemming from, e.g., the $T-$matrix approximation of impurity scattering would be weak, thus validating the choice of an isotropic relaxation time.  Accounting for the anisotropy of the scattering time will provide only quantitative changes to our results. The potential vanishing of $\tilde{\sigma}$ due to the anisotropic corrections $\propto \frac{c}{v_D}$ to the scattering time is unlikely, owing to the accidental origin of such a cancellation. Both statements have been verified numerically. 

We calculate $I^{(1)}_{\lambda {\beta}}, I^{(2,1)}_{\lambda {\beta} {\gamma}}, I^{(2,2)}_{\lambda {\beta} {\gamma}}$ for the model dispersion Eq. \eqref{disp} up to the second order in $\frac{c}{v_D}$ expansion and obtain
\begin{eqnarray}
    &&\sigma_{\lambda \beta} = \frac{e^2}{\frac{1}{\tau} + \alpha(0)} \left[ \frac{\mu}{4 \pi} + \frac{9 \mu}{16 \pi} \left(\frac{c}{v_D} \right)^2 \right] \; \left(1_{2\times2}\right)_{\lambda \beta}, \label{lincondres} \\
    &&\tilde{\sigma}_{\lambda \beta \gamma} = \frac{e^3 c \left[ \frac{5}{\tau} + 11 \alpha(0) \right]}{16 \pi\left(  \frac{1}{\tau} - \alpha(0) \right) \left(  \frac{1}{\tau} + \alpha(0) \right)^2} \left( \delta_{\lambda,x} \tau_3 -\delta_{\lambda,y} \tau_1 \right)_{\beta \gamma}, \nonumber \\ \label{nonlincondres}
\end{eqnarray}
where $\left(1_{2\times2}\right)$ is a $2\times2$ identity matrix, $\delta_{\lambda,x}, \delta_{\lambda,y}$ are Kronecker $\delta-$symbols, and $\tau_{\lambda}$ are Pauli matrices. Both \eqref{lincondres} and \eqref{nonlincondres} respect 3-fold rotational symmetry of the model. We plot linear and nonlinear conductivity tensor prefactors as a function of $\alpha(0)$ in Fig. \ref{fig1}b,c. Expectedly, linear conductivity decreases for $\alpha(0)>0$ (localization) and increases for $\alpha(0)<0$ (antilocalization). The divergence for $\alpha(0)<0$ is unphysical, as for that case $|\alpha(0)| = \frac{1}{\tau}$; hence, physical systems cannot fall into that regime but might approach it. Nonlinear conductivity as a function of $\alpha(0)$ shows an even more interesting behavior: it also experiences an artificial divergence in the antilocalization regime. Importantly, nonlinear conductivity changes sign before hitting the unphysical divergence in the antilocalization regime. The specific value of $\alpha(0)$ at which the sign change occurs depends on specific values of $I^{(2,1)}_{\lambda {\beta} {\gamma}}, I^{(2,2)}_{\lambda {\beta} {\gamma}}$. Nonlinear conductivity also diverges in the localization regime, where linear conductivity is finite and well defined. The sign change of nonlinear conductivity in the localization regime occurs after the singularity is crossed.

To study how $\sigma_{\lambda {\beta}}$ and $\tilde{\sigma}_{\lambda {\beta} {\gamma}}$ depend on density, it is instructive to consider $\tilde{\sigma}_{\lambda {\beta} {\gamma}}/\sigma^2_{\lambda {\beta}}$ as a function of $\alpha(0)$ as for $\alpha(0)=0$ this ratio is independent of the scattering time. The density dependence of $\alpha(0)$ can be estimated from Eq. \eqref{alpha_eq_main}:
\begin{equation}
    \alpha(0) = \pm \frac{\mathcal{C}}{\tau} \frac{4 v_D^4}{\pi \mu \left(4 v_D^2 - 3c^2 \right)} \frac{1}{D} \ln \frac{l_{\phi}}{l}, \label{alpha_estimate_main}
\end{equation}
see Appendix \ref{app:estimation} for details.  In Fig. \ref{fig1}d,e we show the estimated density dependence of linear and nonlinear conductivities for the case of weak localization (blue curves) and antilocalization (orange curves) for our "spinless" model. The unphysical divergence of linear conductivity at low density is again due to $|\alpha(0)| = \frac{1}{\tau}$. Note, however, that in panel e) the sign change of nonlinear conductivity at low density occurs way before the unphysical parameter range is reached.

\section{Experimental relevance and discussion}
\label{sec:experiment_and_conclusions}

To illustrate the main message of this paper, we consider the nonlinear transport in single-layer graphene heterostructures, where inversion symmetry breaking comes from the substrate. This system is chosen for a few reasons. First, there are no observed phase transitions in single-layer graphene in electronic transport; therefore, one doesn't have to account for different electronic ground states, like in Bernal bilayer graphene \cite{Andrea2022,Chichinadze2025GiantNonlinearHall}. Second, in contrast to twisted systems like twisted bilayer graphene, the effects of strain are less pronounced. Hence, one is more justified to assume 3-fold rotational symmetry is intact.

Recently, a systematic extraction procedure for nonlinear conductivity has been developed for disk-shaped samples \cite{Chichinadze2025GiantNonlinearHall}. Since there is no data for linear and nonlinear conductivity measurements in disk geometry for single-layer graphene, we resort to using measurement data in Hall bar geometry \cite{He2022} to compare experiment with theory. In Fig. \ref{fig1}f we show data for Device 4 from the SI of Ref. \cite{He2022}. As is detailed in Appendix \ref{app:data_an}, in the data extraction procedure of Ref. \cite{He2022} $V^{2 \omega}_{xx}/\rho_{xx} \propto \tilde{\sigma}/\sigma^2$. This allows us to compare the data with a theoretical curve for the antilocalization case in Fig. \ref{fig1}e. We observe that extracted $V^{2 \omega}_{xx}/\rho_{xx}$ changes sign as a function of gate voltage and saturates at higher electron density, similar to the behavior of the orange curve in panel e). Such similarity might indicate that weak antilocalization and localization effects could influence nonlinear transport in 2D graphene-based heterostructures.

\textcolor{blue}{}
Within our minimal model nonlinear conductivity $\tilde{\sigma}$ changes sign when 
$\frac{5}{\tau} + 11 \alpha(0) =0,$ as follows from Eq. \eqref{nonlincondres}.
This, together with Eq. \eqref{alpha_estimate_main}, allows us to estimate $\ln \frac{l_{\phi}}{l}:$
$$
\frac{10 \pi^2}{11} D \nu_F = \mathcal{C} \ln \frac{l_\phi}{l}.
$$
We can estimate $D \nu_F$ at the density (or displacement field) for which nonlinear conductivity changes sign in experiment using 
$$
\sigma = \frac{e^2}{\hbar} D \nu_F.
$$
For data from Ref. \cite{Ma2019} nonlinear conductivity changes sign at an electron density $n \simeq 0.5 \times 10^{12} \; \text{cm}^{-2}$, for which linear conductivity $\sigma \simeq 0.4 \frac{e^2}{\hbar}$. This leads to the logarithm estimate to be $\mathcal{C} \left| \ln \frac{l_{\phi}}{l} \right| = 3.6$. Accounting for the spin (the minimal model is spinless) would modify the density of states $\nu_F$ and $\mathcal{C}$. In the "best case scenario" one can estimate $\ln \frac{l_{\phi}}{l} = 1.8$, which is well within the realistic range \cite{Gorbachev2009transition}. For data from Ref. \cite{He2022} nonlinear conductivity sign change appears near $V_g = 20$V, for which $\sigma \simeq 20 \frac{e^2}{\hbar}$. Even after accounting for spin and valley degrees of freedom in $\nu_F$ (i.e., reducing it by a factor of 4 due to spin and valley degeneracy of graphene) would result in, at least, $\left| \ln \frac{l_{\phi}}{l} \right| \sim 45$, which is rather unrealistic. However, this estimate stems from our model-dependent analysis and might change as a result of a more quantitative assessment.

It is important to mention that the effects of weak (anti)localization in single-layer graphene have been identified mostly at higher carrier concentrations  \cite{GrapheneWeakLocalizationSuppressionExperimentPRL2006,GrapheneLocalizationExperiment2008PRL,Gorbachev2009transition,GrapheneRMP}. However, at such densities, there is a delicate interplay between the two effects, both of which are related to trigonal warping. First, nonlinear conductivity is zero for Dirac electrons in the absence of trigonal warping. Therefore, the magnitude of the effect is stimulated by the significance of trigonal warping and is expected to be more pronounced at higher carrier concentrations, where trigonal warping is more prominent. On the other hand, more prominent trigonal warping leads to increased backscattering, which suppresses weak antilocalization \cite{McCannWeakLocalization2006PRL,GrapheneRMP}. Therefore, it is currently not clear whether weak antilocalization effects will be suppressed or pronounced in the nonlinear transport in single-layer graphene devices. 
In this work, we conjectured that in realistic systems that don't experience phase transitions, like single-layer graphene, the possible sign change of nonlinear conductivity as a function of density may be related to weak (anti)localization effects. Our conjecture is based on a simplistic quasiclassical Boltzmann modeling and should be checked with exact microscopic calculations for realistic models, incorporating effects of the scattering rate anisotropy, the substrate, and the perpendicular displacement field, which is the subject of future work.  

\section{Acknowledgments}

The author would like to thank F. Mentink-Vigier, S. Ran, O. Vafek, C. Lewandowski, E. Henriksen, A. Seidel, A. Levchenko, D. Shaffer for fruitful discussions. 
D.V.C. acknowledges financial support from the National High
Magnetic Field Laboratory through a Dirac Fellowship,
which is funded by the National Science Foundation
(Grant No. DMR-2128556) and the State of Florida, and from Washington University in St. Louis through the Edwin Thompson
Jaynes Postdoctoral Fellowship.

\appendix

\section{Boltzmann equation and its solution}


The Boltzmann equation with weak (anti)localization corrections for a distribution function $f(\bp, \br, t)$ in the relaxation time approximation is given by \cite{AmbegaokarPRB1986}
\begin{widetext}
\begin{equation}
    \left( \frac{\partial }{\partial t} + \frac{\partial \br}{\partial t} \cdot \nabla_{\br} + \frac{\partial \bp}{\partial t} \cdot \nabla_{\bp} \right) f(\bp, \br, t) = - \frac{f(\bp, \br, t) - f_{\text{eq}}(\bp, \br, t)}{\tau} + \int_{-\infty}^{t} dt' \; \alpha(t-t') \; \left[ f(-\bp, \br, t') - f_{\text{eq}}(\bp, \br, t') \right],
\end{equation}
where 
\begin{equation}
    \alpha (t-t') = {\pm \frac{\mathcal{C}}{\pi \nu_{F} \tau} \int_{1/l_{\phi}}^{1/l} \frac{d^2 \bq}{(2\pi)^2} e^{-\left( D \bq^2  \right) \left(t-t' \right)}}.
\end{equation}
The $+$ sign in the definition above corresponds to the weak localization case, whereas $-$ sign appears for weak antilocalization \cite{KimPRB2014}; $\nu_F$ -- is the density of states (DOS) at the Fermi level; the upper and lower cutoffs are given by mean-free path and the phase-coherence length correspondingly, $D$ -- is the diffusion constant, and $\mathcal{C}$ -- is a numerical prefactor that accounts for spin and/or isospin structure of the system. The equilibrium distribution function $f_{\text{eq}}(\bp, \br, t)$ is time-independent but we keep the redundant time index for the sake of bookkeeping. We also set $\hbar = 1$ throughout this calculation. 

In this work we consider a spatially homogeneous case with $f(\bp, \br, t) = f(\bp, t)$. For the applied external electric field $\dot{\bp} = e \bE$, thus, the Boltzmann equation takes the form
\begin{equation}
    \left( \frac{\partial }{\partial t} + e \bE \cdot \nabla_{\bp} \right) f(\bp, t) = - \frac{f(\bp, t) - f_{\text{eq}}(\bp, t)}{\tau} + \int_{-\infty}^{t} dt' \; \alpha(t-t') \; \left[ f(-\bp, t') - f_{\text{eq}}(\bp, t') \right].
    \label{Boltzmann_time}
\end{equation}
It is more convenient to solve equations like Eq. \eqref{Boltzmann_time} in frequency domain, therefore, we introduce 
$$
f (\bp, t) = \int d \omega' e^{- i \omega' t} f (\bp, \omega')
$$
and substitute it into Eq. \eqref{Boltzmann_time}:
\begin{equation}
\begin{gathered}
    \int d \omega' e^{- i \omega' t} \left( - i \omega' + e \bE \cdot \nabla_{\bp} \right) f(\bp, \omega') = \\ = - \int d \omega' e^{- i \omega' t} \frac{f(\bp, \omega') - f_{\text{eq}}(\bp, \omega')}{\tau} + \int d \omega' e^{- i \omega' t}  \left[ f(-\bp, \omega') - f_{\text{eq}}(\bp, \omega') \right] \int_{-\infty}^{t} dt' e^{i \omega' \left( t - t' \right)} \; \alpha(t-t').
    \end{gathered}
    \label{Boltzmann_intermediate}
\end{equation}
We introduce a "partial" Fourier transform 
\begin{equation}
    \alpha (\omega') = \int_{-\infty}^{t} dt' e^{i \omega' \left( t - t' \right)} \; \alpha(t-t'),
\end{equation}
multiply both sides of Eq. \eqref{Boltzmann_intermediate} by $\frac{e^{i \omega t}}{2\pi}$, and integrate the whole expression over $t$ from $-\infty$ to $\infty$ to obtain the frequency domain representation of the Boltzmann equation
\begin{equation}
    \left( - i \omega + e \bE \cdot \nabla_{\bp} \right) f(\bp, \omega) = -  \frac{f(\bp, \omega) - f_{\text{eq}}(\bp, \omega)}{\tau} + \alpha(\omega) \left[ f(-\bp, \omega) - f_{\text{eq}}(\bp, \omega) \right].
    \label{Boltzmann}
\end{equation}
The function $\alpha(\omega)$ can be calculated explicitly
\begin{equation}
    \begin{gathered}
        \alpha (\omega) = \int_{-\infty}^{t} dt' e^{i \omega \left( t - t' \right)} \; \alpha(t-t') = \int_{-\infty}^{t} dt' e^{i \omega \left( t - t' \right)} \; {\left( \pm \frac{\mathcal{C}}{\pi \nu_{F} \tau} \int_{1/l_{\phi}}^{1/l} \frac{d^2 \bq}{(2\pi)^2} e^{-\left( D \bq^2  \right) \left(t-t' \right)} \right)} = \\
        = {\pm \frac{\mathcal{C}}{\pi \nu_{F} \tau} \int_{1/l_{\phi}}^{1/l} \frac{d^2 \bq}{(2\pi)^2} \int_{-\infty}^{t} dt' \; e^{-\left( D \bq^2 - i \omega  \right) \left(t-t' \right)} = \pm \frac{\mathcal{C}}{\pi \nu_{F} \tau} \int_{1/l_{\phi}}^{1/l} \frac{d^2 \bq}{(2\pi)^2} \frac{1}{D \bq^2 - i \omega }},
    \end{gathered}
    \label{alpha_eq}
\end{equation}
if $D \bq^2 >0,$ which is manifestly true within the integration range and renders this standard weak (anti)localization theory integral being positive definite for $\omega=0$ \cite{AmbegaokarPRB1986}:
\begin{equation}
    \begin{gathered}
        {\int_{1/l_{\phi}}^{1/l} \frac{d^2 \bq}{(2\pi)^2} \frac{1}{D \bq^2 } = \frac{1}{2 \pi D} \ln \frac{l_{\phi}}{l}},
    \end{gathered}
\end{equation}
where $l$ - is the mean-free path and $l_{\phi}$ is the phase-coherence length.

We now proceed to solving Eq. \eqref{Boltzmann} up to the second order in the external electric field $\bE$ by expanding 
$$
f(\bp, \omega) = f_0(\bp, \omega) + f_1(\bp, \omega) + f_2(\bp, \omega)+ ...
$$
where $f_i(\bp, \omega) \sim (\bE)^i$. We make two important remarks here. First, in this approach $f_{\text{eq}}(\bp, \omega) = \frac{1}{e^{{\frac{\varepsilon_\bp - \mu}{T}}}+1} \delta (\omega)$ is given by a (frequency-independent) Fermi function because the equilibrium distribution function is time-independent. Therefore, even though the formal equilibrium solution of Eq. \eqref{Boltzmann} for finite $\omega$ is given by a frequency-dependent $f_0(\bp, \omega)$, the $\delta$-function in $f_{\text{eq}}(\bp, \omega)$ effectively leads to $f_0(\bp, \omega)=f_{\text{eq}}(\bp, \omega) = f_{\text{eq}}(\bp, 0)$, as we'll see below.
Second, in our solutions we \underline{do not} assume $f(\bp, \omega) = f(-\bp, \omega)$ symmetry of the distribution function.

\subsection{Solution to order $\bE^0$}

We first solve for the distribution function $f_0(\bp, \omega),$ which is independent of the electric field. The Boltzmann equation at 0th order in $\bE$ reads
\begin{eqnarray}
     - i \omega f_0(\bp, \omega) &=& -  \frac{f_0(\bp, \omega) - f_{\text{eq}}(\bp, \omega)}{\tau} + \alpha(\omega) \left[ f_0(-\bp, \omega) - f_{\text{eq}}(\bp, \omega) \right], \\
     - i \omega f_0(-\bp, \omega) &=& -  \frac{f_0(-\bp, \omega) - f_{\text{eq}}(-\bp, \omega)}{\tau} + \alpha(\omega) \left[ f_0(\bp, \omega) - f_{\text{eq}}(-\bp, \omega) \right],
\end{eqnarray}
where the second equation was obtained using a $\bp \rightarrow -\bp$ transformation. 
The two equations are to be considered on equal grounds as, in general, $f_0(\bp, \omega) \neq f_0(-\bp, \omega)$ and $f_{\text{eq}}(\bp, \omega) \neq f_{\text{eq}}(-\bp, \omega)$. By adding and subtracting the two equations we arrive at 
\begin{eqnarray}
    f_0(\bp, \omega) + f_0(-\bp, \omega) &=& \frac{ \frac{1}{\tau} - \alpha(\omega) }{- i \omega + \frac{1}{\tau} - \alpha(\omega)} \left[ f_{\text{eq}}(\bp, \omega) + f_{\text{eq}}(-\bp, \omega) \right], \label{Boltzmannf0} \\
    f_0(\bp, \omega) - f_0(-\bp, \omega) &=& \frac{ \frac{1}{\tau} - \alpha(\omega) }{- i \omega + \frac{1}{\tau} + \alpha(\omega)} \left[ f_{\text{eq}}(\bp, \omega) - f_{\text{eq}}(-\bp, \omega) \right],
\end{eqnarray}
which immediately yields the solution for $f_0(\bp, \omega)$:
\begin{equation}
    f_0(\bp, \omega) = \frac{1}{2} \left[ \frac{ \frac{1}{\tau} - \alpha(\omega) }{- i \omega + \frac{1}{\tau} - \alpha(\omega)} \left[ f_{\text{eq}}(\bp, \omega) + f_{\text{eq}}(-\bp, \omega) \right] + \frac{ \frac{1}{\tau} - \alpha(\omega) }{- i \omega + \frac{1}{\tau} + \alpha(\omega)} \left[ f_{\text{eq}}(\bp, \omega) - f_{\text{eq}}(-\bp, \omega) \right] \right].
    \label{f0sol}
\end{equation}
One can check by a direct substitution that \eqref{f0sol} solves Eq. \eqref{Boltzmannf0}. 
Imposing $f_{\text{eq}}(\bp, \omega) = f_{\text{eq}}(-\bp, \omega)$ we get
\begin{equation}
    f_0(\bp, \omega) = \frac{ \frac{1}{\tau} - \alpha(\omega) }{- i \omega + \frac{1}{\tau} - \alpha(\omega)} f_{\text{eq}}(\bp, \omega).
\end{equation}
Given that within our approach the equilibrium distribution function has meaning only for zero frequency, we set $\omega=0$ and recover the expected relation 
\begin{equation}
    f_0(\bp, 0) = f_{\text{eq}}(\bp, 0).
\end{equation}

\subsection{Solution to order $\bE^1$}

At the linear order in $\bE$ the pair of equations we ought to solve is 
\begin{eqnarray}
     - i \omega f_1(\bp, \omega) + e \bE \cdot \nabla_{\bp} f_0(\bp, \omega) &=& -  \frac{f_1(\bp, \omega)}{\tau} + \alpha(\omega) f_1(-\bp, \omega), \\
     - i \omega f_1(-\bp, \omega) - e \bE \cdot \nabla_{\bp} f_0(-\bp, \omega) &=& -  \frac{f_1(-\bp, \omega)}{\tau} + \alpha(\omega) f_1(\bp, \omega),
\end{eqnarray}
where the second equation was again obtained from the first one with the $\bp \rightarrow - \bp$ transformation. Note the opposite signs in front of the gradient terms in both equations. Adding and subtracting the two equations leads to
\begin{eqnarray}
    f_1(\bp, \omega) + f_1(-\bp, \omega) &=& - \frac{e \bE \cdot \nabla_{\bp}}{-i \omega + \frac{1}{\tau} - \alpha(\omega)} \left[ f_0(\bp, \omega) - f_0(-\bp, \omega) \right], \\
    f_1(\bp, \omega) - f_1(-\bp, \omega) &=& - \frac{e \bE \cdot \nabla_{\bp}}{-i \omega + \frac{1}{\tau} + \alpha(\omega)} \left[ f_0(\bp, \omega) + f_0(-\bp, \omega) \right],
\end{eqnarray}
and, therefore,
\begin{equation}
    f_1(\bp, \omega) = - \frac{e \bE \cdot \nabla_{\bp}}{2} \left[ \frac{f_0(\bp, \omega) - f_0(-\bp, \omega)}{-i \omega + \frac{1}{\tau} - \alpha(\omega)} + \frac{f_0(\bp, \omega) + f_0(-\bp, \omega)}{-i \omega + \frac{1}{\tau} + \alpha(\omega)} \right].
    \label{f1sol}
\end{equation}
Importantly, the expression for the next-order distribution function will retain the form of  Eq. \eqref{f1sol}.

Substituting expressions for $f_0(\bp, \omega)$ and $f_0(-\bp, \omega)$ into Eq. \eqref{f1sol} and taking the limit of $\omega \rightarrow 0$ we arrive at the well-known expression
\begin{equation}
    f_1(\bp, \omega) = - \frac{e \bE \cdot \nabla_{\bp}}{\frac{1}{\tau} + \alpha(0)} f_{\text{eq}}(\bp).
\end{equation}
Note that for weak localization $\alpha(0)>0$, which increases scattering rate and, thus, reduces linear conductivity, as expected.

\subsection{Solution to order $\bE^2$}

Finally, we solve for the distribution function $f_2(\bp, \omega)$. At the 2nd order in $\bE$, the pair of equations that define the distribution function is
\begin{eqnarray}
     - i \omega f_2(\bp, \omega) + e \bE \cdot \nabla_{\bp} f_1(\bp, \omega) &=& -  \frac{f_2(\bp, \omega)}{\tau} + \alpha(\omega) f_2(-\bp, \omega), \\
     - i \omega f_2(-\bp, \omega) - e \bE \cdot \nabla_{\bp} f_1(-\bp, \omega) &=& -  \frac{f_2(-\bp, \omega)}{\tau} + \alpha(\omega) f_2(\bp, \omega),
\end{eqnarray}
which has the same structure as in the linear in $\bE$ case. 
Performing exactly the same manipulations as at the order $\bE$ we get 
\begin{eqnarray}
    f_2(\bp, \omega) + f_2(-\bp, \omega) &=& - \frac{e \bE \cdot \nabla_{\bp}}{-i \omega + \frac{1}{\tau} - \alpha(\omega)} \left[ f_1(\bp, \omega) - f_1(-\bp, \omega) \right], \\
    f_2(\bp, \omega) - f_2(-\bp, \omega) &=& - \frac{e \bE \cdot \nabla_{\bp}}{-i \omega + \frac{1}{\tau} + \alpha(\omega)} \left[ f_1(\bp, \omega) + f_1(-\bp, \omega) \right],
\end{eqnarray}
and the solution reads
\begin{equation}
    f_2(\bp, \omega) = - \frac{e \bE \cdot \nabla_{\bp}}{2} \left[ \frac{f_1(\bp, \omega) - f_1(-\bp, \omega)}{-i \omega + \frac{1}{\tau} - \alpha(\omega)} + \frac{f_1(\bp, \omega) + f_1(-\bp, \omega)}{-i \omega + \frac{1}{\tau} + \alpha(\omega)} \right].
    \label{f2sol}
\end{equation}

Substituting solutions for $f_1 (\bp, \omega)$ and taking the limit of $\omega \rightarrow 0$ we get
\begin{equation}
    f_2(\bp, 0) = - \frac{e^2 E_{{\beta}} E_{{\gamma}} \frac{\partial}{\partial p_{{\beta}}} \frac{\partial}{\partial p_{{\gamma}}} \left[ \frac{1}{\tau} f_{\text{eq}}(\bp) + \alpha(0) f_{\text{eq}}(-\bp) \right] }{\left(  -\frac{1}{\tau} + \alpha(0) \right) \left(  \frac{1}{\tau} + \alpha(0) \right)^2}. 
\end{equation}
Assuming independence of $\alpha(0)$ and $\tau$ one can formally take a "quasiballistic" limit of $\tau \rightarrow \infty$ and obtain 
\begin{equation}
    f_2(\bp, 0) = - \frac{e^2 E_{{\beta}} E_{{\gamma}} \frac{\partial}{\partial p_{{\beta}}} \frac{\partial}{\partial p_{{\gamma}}} \left[ f_{\text{eq}}(-\bp) \right] }{ \left(  \alpha(0) \right)^2}. 
\end{equation}
In this somewhat artificial case, we end up with a peculiar result that for both signs of $\alpha(0)$ the resulting distribution function and, therefore, nonlinear conductivity tensor, is given by the same expression. It could indicate that weak localization and antilocalization effects in an extremely clean system might give the same nonlinear conductivity tensor.

\section{Linear and nonlinear conductivity tensors}

To calculate linear and nonlinear conductivity tensors in DC limit we employ the definition of current density
\begin{equation}
    j_{\lambda} = e \int \frac{d^2 \bp}{(2 \pi)^2} v_{\lambda} f(\bp),
\end{equation}
where $v_{\lambda} = \frac{\partial \varepsilon_{\bp}}{\partial p_{\lambda}}$ is a component of the charge carriers' velocity, $\varepsilon_{\bp}$ is the quasiparticle dispersion, and $f(\bp)$ is the distribution function.

\subsection{Linear conductivity tensor}

Linear conductivity tensor relates current density $j_{\lambda}$ and linear in $\bE$ contribution to distribution function:
\begin{equation}
    \begin{gathered}
        j_{\lambda} = e \int \frac{d^2 \bp}{(2 \pi)^2} v_{\lambda} f_1(\bp) = -\frac{e^2}{\frac{1}{\tau} + \alpha(0)} \int \frac{d^2 \bp}{(2 \pi)^2} v_{\lambda} \left[\bE \cdot \nabla_{\bp} f_{\text{eq}}(\bp) \right] = -\frac{e^2 E_{{\beta}}}{\frac{1}{\tau} + \alpha(0)} \int \frac{d^2 \bp}{(2 \pi)^2} v_{\lambda} \frac{\partial f_{\text{eq}}(\bp)}{\partial p_{{\beta}}}, 
    \end{gathered}
\end{equation}
which allows us to introduce a linear conductivity tensor
\begin{equation}
    \sigma_{\lambda {\beta}} = -\frac{e^2}{\frac{1}{\tau} + \alpha(0)} \int \frac{d^2 \bp}{(2 \pi)^2} v_{\lambda} \frac{\partial f_{\text{eq}}(\bp)}{\partial p_{{\beta}}} = -\frac{e^2}{\frac{1}{\tau} + \alpha(0)} I^{(1)}_{\lambda {\beta}},
\end{equation}
such that $j_{\lambda} = \sigma_{\lambda {\beta}} E_{{\beta}}$ and
\begin{equation}
    I^{(1)}_{\lambda {\beta}} = \int \frac{d^2 \bp}{(2 \pi)^2} v_{\lambda} \frac{\partial f_{\text{eq}}(\bp)}{\partial p_{{\beta}}} = \int \frac{d^2 \bp}{(2 \pi)^2} \frac{\partial \varepsilon_{\bp}}{\partial p_{\lambda}} \frac{\partial \varepsilon_{\bp}}{\partial p_{{\beta}}} \frac{\partial f_{\text{eq}}(\varepsilon)}{\partial \varepsilon}.
\end{equation}

\subsection{Nonlinear conductivity tensor}

Quadratic in $\bE$ contribution to current density in the DC limit reads
\begin{equation}
    \begin{gathered}
        j_{\lambda} = e \int \frac{d^2 \bp}{(2 \pi)^2} v_{\lambda} f_2(\bp) = e \int \frac{d^2 \bp}{(2 \pi)^2} v_{\lambda} \left[ - \frac{e^2 E_{{\beta}} E_{{\gamma}} \frac{\partial}{\partial p_{{\beta}}} \frac{\partial}{\partial p_{{\gamma}}} \left[ \frac{1}{\tau} f_{\text{eq}}(\bp) + \alpha(0) f_{\text{eq}}(-\bp) \right] }{\left(  -\frac{1}{\tau} + \alpha(0) \right) \left(  \frac{1}{\tau} + \alpha(0) \right)^2} \right] = \\
        = - \frac{e^3 E_{{\beta}} E_{{\gamma}}}{\left(  -\frac{1}{\tau} + \alpha(0) \right) \left(  \frac{1}{\tau} + \alpha(0) \right)^2} \int \frac{d^2 \bp}{(2 \pi)^2} v_{\lambda} \left[  \frac{\partial}{\partial p_{{\beta}}} \frac{\partial}{\partial p_{{\gamma}}} \left[ \frac{1}{\tau} f_{\text{eq}}(\bp) + \alpha(0) f_{\text{eq}}(-\bp) \right]  \right] = \\
        = \frac{e^3 E_{{\beta}} E_{{\gamma}}}{\left(  \frac{1}{\tau} - \alpha(0) \right) \left(  \frac{1}{\tau} + \alpha(0) \right)^2} \left[ \frac{1}{\tau} I^{(2,1)}_{\lambda {\beta} {\gamma}} + \alpha(0) I^{(2,2)}_{\lambda {\beta} {\gamma}} \right] = \tilde{\sigma}_{\lambda {\beta} {\gamma}} E_{{\beta}} E_{{\gamma}},
    \end{gathered}
\end{equation}
where we introduced two model-specific integrals
\begin{eqnarray}
    I^{(2,1)}_{\lambda {\beta} {\gamma}} &=& \int \frac{d^2 \bp}{(2 \pi)^2} \frac{\partial \varepsilon_{\bp}}{\partial p_{\lambda}}   \frac{\partial^2 f_{\text{eq}}(\bp)}{\partial p_{{\beta}} \partial p_{{\gamma}}}, \\
    I^{(2,2)}_{\lambda {\beta} {\gamma}} &=& \int \frac{d^2 \bp}{(2 \pi)^2} \frac{\partial \varepsilon_{\bp}}{\partial p_{\lambda}}  \frac{\partial^2 f_{\text{eq}}(-\bp)}{\partial p_{{\beta}} \partial p_{{\gamma}}}
\end{eqnarray}
and the nonlinear conductivity tensor
\begin{equation}
    \tilde{\sigma}_{\lambda {\beta} {\gamma}} = \frac{e^3}{\left(  \frac{1}{\tau} - \alpha(0) \right) \left(  \frac{1}{\tau} + \alpha(0) \right)^2} \left[ \frac{1}{\tau} I^{(2,1)}_{\lambda {\beta} {\gamma}} + \alpha(0) I^{(2,2)}_{\lambda {\beta} {\gamma}} \right].
\end{equation}
Using the chain rule and integration by parts the expressions for the two integrals can be cast into the following form:
\begin{eqnarray}
    I^{(2,1)}_{\lambda {\beta} {\gamma}} &=& \left( \frac{\partial \varepsilon(\bp)}{\partial p_{\lambda}} \frac{\partial \varepsilon(\bp)}{\partial p_{{\gamma}}} \frac{\partial f_{\text{eq}} (\varepsilon)}{\partial \varepsilon} \right) \Biggr|_{\Gamma} - \int \frac{d^2 \bp}{(2 \pi)^2} \left( \frac{\partial f_{\text{eq}} (\varepsilon)}{\partial \varepsilon} \right) \left( \frac{\partial^2 \varepsilon(\bp)}{\partial p_{\lambda} \partial p_{{\beta}} } \right) \left( \frac{\partial \varepsilon(\bp)}{\partial p_{{\gamma}}} \right), \label{I21integral} \\
    I^{(2,2)}_{\lambda {\beta} {\gamma}} &=& \left( \frac{\partial \varepsilon(\bp)}{\partial p_{\lambda}} \frac{\partial \varepsilon(-\bp)}{\partial p_{{\gamma}}} \frac{\partial f_{\text{eq}} (\varepsilon)}{\partial \varepsilon} \right) \Biggr|_{\Gamma} - \int \frac{d^2 \bp}{(2 \pi)^2} \left( \frac{\partial f_{\text{eq}} (\varepsilon)}{\partial \varepsilon} \right) \left( \frac{\partial^2 \varepsilon(\bp)}{\partial p_{\lambda} \partial p_{{\beta}} } \right) \left( \frac{\partial \varepsilon(-\bp)}{\partial p_{{\gamma}}} \right), \label{I22integral}
\end{eqnarray}
where $\Gamma$ is the boundary of the integration region.

\section{Calculation for a model dispersion}

In this work we consider a low energy model dispersion which describes a free Dirac electron in the presence of trigonal warping with the dispersion given by
$$
\varepsilon_{\bk} = v_D |\bk| + c |\bk| \cos 3\theta_{\bk} = v_D \sqrt{k_x^2 + k_y^2} + c \left( \frac{k_x^3 - 3 k_x k_y^2}{k_x^2 + k_y^2} \right).
$$
Here $v_D$ is the Dirac velocity and $c$ controls the strength of trigonal warping. We use this model as a proxy for a realistic model of graphene and omit all Berry curvature effects, as they are out of scope of this work. 
In our further calculations we will find useful the following expressions for derivatives of the dispersion
\begin{eqnarray}
    \frac{\partial \varepsilon_{\bp}}{\partial p_{x}} &=& v_D \cos \theta + 2 c \cos 2 \theta - c \cos 4 \theta = v_D \left( \cos \theta + 2 \frac{c}{v_D} \cos 2 \theta - \frac{c}{v_D} \cos 4 \theta \right), \nonumber \\ \nonumber
    \frac{\partial \varepsilon_{\bp}}{\partial p_{y}} &=& v_D \sin \theta - 2 c \sin 2 \theta - c \sin 4 \theta = v_D \left( \sin \theta - 2 \frac{c}{v_D} \sin 2 \theta - \frac{c}{v_D} \sin 4 \theta \right), \\ \nonumber
    \frac{\partial \varepsilon_{-\bp}}{\partial p_{x}} &=& v_D \cos \theta - 2 c \cos 2 \theta + c \cos 4 \theta = v_D \left( \cos \theta - 2 \frac{c}{v_D} \cos 2 \theta + \frac{c}{v_D} \cos 4 \theta \right), \\ \nonumber
    \frac{\partial \varepsilon_{-\bp}}{\partial p_{y}} &=& v_D \sin \theta + 2 c \sin 2 \theta + c \sin 4 \theta = v_D \left( \sin \theta + 2 \frac{c}{v_D} \sin 2 \theta + \frac{c}{v_D} \sin 4 \theta \right), \\ \nonumber
    \frac{\partial^2 \varepsilon_{\bp}}{\partial p_{x}^2} &=& \frac{\sin^2 \theta}{k} \left( v_D - 8 c \cos 3 \theta \right) =  v_D \frac{1- \cos 2\theta}{2k} \left( 1 - 8 \frac{c}{v_D} \cos 3 \theta \right), \\ \nonumber
    \frac{\partial^2 \varepsilon_{\bp}}{\partial p_{x} \partial p_{y}} &=& - \frac{\sin 2 \theta}{2k} \left( v_D - 8 c \cos 3 \theta \right) =  - v_D \frac{\sin 2 \theta}{2k} \left( 1 - 8 \frac{c}{v_D} \cos 3 \theta \right), \\ \nonumber
    \frac{\partial^2 \varepsilon_{\bp}}{\partial p_{y}^2} &=& \frac{\cos^2 \theta}{k} \left( v_D - 8 c \cos 3 \theta \right) =  v_D \frac{1 + \cos 2\theta}{2k} \left( 1 - 8 \frac{c}{v_D} \cos 3 \theta \right).
\end{eqnarray}

\subsection{Density of states}

At the Fermi level electronic density of states (DOS) is given by
\begin{equation}
    \nu_F = \int \frac{d^2 \bp}{(2 \pi)^2} \delta ( \varepsilon_{\bp} - \mu ) = \int \frac{d \theta}{(2 \pi)^2} \frac{k(\theta)}{\left| \bv (k(\theta)) \right|}, 
\end{equation}
where the Fermi surface and Fermi velocity for chemical potential $\mu$ are given by
\begin{eqnarray}
        |\bk| &=& k(\theta) = \frac{\mu}{v_D + c \cos 3\theta_{\bk}}, \nonumber \\
    \left| \nabla_{\bp} \varepsilon_{\bp} \right| = \left| \bv(\bk) \right| &=& \sqrt{5 c^2 + v_D^2 + 2 c v_D \cos 3 \theta - 4 c^2 \cos 6 \theta}. \nonumber
\end{eqnarray}
Expanding the expression up to the second order in $\frac{c}{v_D}$ and taking the integral we get
\begin{equation}
    \nu_F = \mu \frac{4 v_D^2 - 3c^2}{8 \pi v_D^4}.
    \label{eq_dos}
\end{equation}

\subsection{Linear conductivity tensor}

We assume electron doping of the system; in such case $\mu>0$ and one can safely ignore the boundary term as the states at the high momentum cutoff are empty. 
Working at $T=0$ the integral $I^{(1)}_{\lambda {\beta}}$ can be rewritten as
\begin{eqnarray}
        I^{(1)}_{\lambda {\beta}} &=& \int \frac{d^2 \bp}{(2 \pi)^2} \frac{\partial \varepsilon_{\bp}}{\partial p_{\lambda}} \frac{\partial \varepsilon_{\bp}}{\partial p_{{\beta}}} \frac{\partial f_{\text{eq}}(\varepsilon)}{\partial \varepsilon} = - \int \frac{d^2 \bp}{(2 \pi)^2} \frac{\partial \varepsilon_{\bp}}{\partial p_{\lambda}} \frac{\partial \varepsilon_{\bp}}{\partial p_{{\beta}}} \delta ( \varepsilon_{\bp} - \mu ) = - \int \frac{d \theta}{(2 \pi)^2} \frac{k(\theta)}{\left| \bv (k(\theta)) \right|} \frac{\partial \varepsilon_{\bp}}{\partial p_{\lambda}} \frac{\partial \varepsilon_{\bp}}{\partial p_{{\beta}}}.
\end{eqnarray}
Making use of the expressions above we calculate
\begin{eqnarray}
        I^{(1)}_{xx} &=& - \int \frac{d \theta}{(2 \pi)^2} \frac{\mu}{v_D} \frac{1}{1 + \frac{c}{v_D} \cos 3\theta_{\bk}} \frac{1}{v_D \sqrt{5 \left( \frac{c}{v_D} \right)^2 + 1 + 2 \frac{c}{v_D} \cos 3 \theta - 4 \left( \frac{c}{v_D} \right)^2 \cos 6 \theta}} \frac{\partial \varepsilon_{\bp}}{\partial p_{x}} \frac{\partial \varepsilon_{\bp}}{\partial p_{x}} \nonumber \\
        &=& - \int \frac{d \theta}{(2 \pi)^2} \frac{\mu}{1 + \frac{c}{v_D} \cos 3\theta_{\bk}} \frac{\left( \cos \theta + 2 \frac{c}{v_D} \cos 2 \theta - \frac{c}{v_D} \cos 4 \theta \right)^2}{\sqrt{5 \left( \frac{c}{v_D} \right)^2 + 1 + 2 \frac{c}{v_D} \cos 3 \theta - 4 \left( \frac{c}{v_D} \right)^2 \cos 6 \theta}} \simeq - \frac{\mu}{4 \pi} - \frac{9 \mu}{16 \pi} \left( \frac{c}{v_D} \right)^2 
\end{eqnarray}
and
\begin{eqnarray}
        I^{(1)}_{yy} &=& - \int \frac{d \theta}{(2 \pi)^2} \frac{\mu}{v_D} \frac{1}{1 + \frac{c}{v_D} \cos 3\theta_{\bk}} \frac{1}{v_D \sqrt{5 \left( \frac{c}{v_D} \right)^2 + 1 + 2 \frac{c}{v_D} \cos 3 \theta - 4 \left( \frac{c}{v_D} \right)^2 \cos 6 \theta}} \frac{\partial \varepsilon_{\bp}}{\partial p_{y}} \frac{\partial \varepsilon_{\bp}}{\partial p_{y}} \nonumber \\
        &=& - \int \frac{d \theta}{(2 \pi)^2} \frac{\mu}{1 + \frac{c}{v_D} \cos 3\theta_{\bk}} \frac{\left( \sin \theta - 2 \frac{c}{v_D} \sin 2 \theta - \frac{c}{v_D} \sin 4 \theta \right)^2}{\sqrt{5 \left( \frac{c}{v_D} \right)^2 + 1 + 2 \frac{c}{v_D} \cos 3 \theta - 4 \left( \frac{c}{v_D} \right)^2 \cos 6 \theta}} \simeq - \frac{\mu}{4 \pi} - \frac{9 \mu}{16 \pi} \left(\frac{c}{v_D} \right)^2
\end{eqnarray}
up to second order in $\frac{c}{v_D}$. For off-diagonal components $I^{(1)}_{xy}$ and $I^{(1)}_{yx}$ we get 0 in the similar way. Hence, for our model dispersion the linear conductivity tensor reads
\begin{equation}
    \sigma_{\lambda {\beta}} = -\frac{e^2}{\frac{1}{\tau} + \alpha(0)} I^{(1)}_{\lambda {\beta}} = \frac{e^2}{\frac{1}{\tau} + \alpha(0)} \left[ \frac{\mu}{4 \pi} + \frac{9 \mu}{16 \pi} \left(\frac{c}{v_D} \right)^2 \right] \; \left(1_{2\times2}\right)_{\lambda {\beta}},
    \label{lin_res}
\end{equation}
where $1_{2\times2}$ is a 2-by-2 identity matrix.

\subsection{Nonlinear conductivity tensor}

As in the case of linear conductivity calculation, we assume electron doping and safely neglect the boundary term. In such case, the two integrals we're after are
\begin{eqnarray}
    I^{(2,1)}_{\lambda {\beta} {\gamma}} &=& - \int \frac{d^2 \bp}{(2 \pi)^2} \left( \frac{\partial f_{\text{eq}} (\varepsilon)}{\partial \varepsilon} \right) \left( \frac{\partial^2 \varepsilon(\bp)}{\partial p_{\lambda} \partial p_{{\beta}}} \right) \left( \frac{\partial \varepsilon(\bp)}{\partial p_{{\gamma}}} \right) = \int \frac{d^2 \bp}{(2 \pi)^2} \left( \frac{\partial^2 \varepsilon(\bp)}{\partial p_{\lambda} \partial p_{{\beta}}} \right) \left( \frac{\partial \varepsilon(\bp)}{\partial p_{{\gamma}}} \right) \delta (\varepsilon_{\bp} - \mu) = \nonumber \\
    &=& \int \frac{d \theta}{(2 \pi)^2} \frac{k(\theta)}{\left| \bv (k (\theta)) \right|} \left( \frac{\partial^2 \varepsilon(\bp)}{\partial p_{\lambda} \partial p_{{\beta}}} \right) \left( \frac{\partial \varepsilon(\bp)}{\partial p_{{\gamma}}} \right), \nonumber \\
    I^{(2,2)}_{\lambda {\beta} {\gamma}} &=& - \int \frac{d^2 \bp}{(2 \pi)^2} \left( \frac{\partial f_{\text{eq}} (\varepsilon)}{\partial \varepsilon} \right) \left( \frac{\partial^2 \varepsilon(\bp)}{\partial p_{\lambda} \partial p_{{\beta}}} \right) \left( \frac{\partial \varepsilon(-\bp)}{\partial p_{{\gamma}}} \right) = \int \frac{d^2 \bp}{(2 \pi)^2} \left( \frac{\partial^2 \varepsilon(\bp)}{\partial p_{\lambda} \partial p_{{\beta}}} \right) \left( \frac{\partial \varepsilon(-\bp)}{\partial p_{{\gamma}}} \right) \delta ( \varepsilon_{-\bp} -  \mu ) = \nonumber \\
    &=& \int \frac{d \theta}{(2 \pi)^2} \frac{k(\theta)}{\left| \nabla_{\bp} \varepsilon_{-\bp}  (k (\theta)) \right|} \left( \frac{\partial^2 \varepsilon(\bp)}{\partial p_{\lambda} \partial p_{{\beta}}} \right) \left( \frac{\partial \varepsilon(-\bp)}{\partial p_{{\gamma}}} \right). \nonumber
\end{eqnarray}
Note, that in the second integral the integration contour and DOS are given by
\begin{eqnarray}
        |\bk| &=& k(\theta) = \frac{\mu}{v_D - c \cos 3\theta_{\bk}}, \nonumber \\
    \left| \nabla_{\bp} \varepsilon_{-\bp} \right| &=& \sqrt{5 c^2 + v_D^2 - 2 c v_D \cos 3 \theta - 4 c^2 \cos 6 \theta}. \nonumber
\end{eqnarray}

Using the expressions from the beginning of this section and the parameterization for the Fermi surface we can calculate the first integral using the expansion up to the second order in $\frac{c}{v_D}$:
\begin{eqnarray}
    I^{(2,1)}_{xxx} &=& \int \frac{d \theta}{(2 \pi)^2} \frac{\mu}{v_D} \frac{1}{1 + \frac{c}{v_D} \cos 3\theta_{\bk}} \frac{1}{v_D \sqrt{5 \left( \frac{c}{v_D} \right)^2 + 1 + 2 \frac{c}{v_D} \cos 3 \theta - 4 \left( \frac{c}{v_D} \right)^2 \cos 6 \theta}} \left( \frac{\partial^2 \varepsilon(\bp)}{\partial p_{x} \partial p_{x}} \right) \left( \frac{\partial \varepsilon(\bp)}{\partial p_{x}} \right) = \nonumber \\
    &=& \int \frac{d \theta}{(2 \pi)^2} \frac{v_D (1- \cos 2\theta) \left( 1 - 8 \frac{c}{v_D} \cos 3 \theta \right) \left( \cos \theta + 2 \frac{c}{v_D} \cos 2 \theta - \frac{c}{v_D} \cos 4 \theta \right)}{2  \sqrt{5 \left( \frac{c}{v_D} \right)^2 + 1 + 2 \frac{c}{v_D} \cos 3 \theta - 4 \left( \frac{c}{v_D} \right)^2 \cos 6 \theta}} \simeq \frac{5 c}{16 \pi}.
\end{eqnarray}
The other non-zero terms have the same magnitude but the opposite sign and, when calculated using the expansion up to second order in $\frac{c}{v_D}$, read
\begin{eqnarray}
    I^{(2,1)}_{yyx} &=& \int \frac{d \theta}{(2 \pi)^2} \frac{\mu}{v_D} \frac{1}{1 + \frac{c}{v_D} \cos 3\theta_{\bk}} \frac{1}{v_D \sqrt{5 \left( \frac{c}{v_D} \right)^2 + 1 + 2 \frac{c}{v_D} \cos 3 \theta - 4 \left( \frac{c}{v_D} \right)^2 \cos 6 \theta}} \left( \frac{\partial^2 \varepsilon(\bp)}{\partial p_{y} \partial p_{y}} \right) \left( \frac{\partial \varepsilon(\bp)}{\partial p_{x}} \right) = \nonumber \\
    &=& \int \frac{d \theta}{(2 \pi)^2} \frac{v_D (1+ \cos 2\theta) \left( 1 - 8 \frac{c}{v_D} \cos 3 \theta \right) \left( \cos \theta + 2 \frac{c}{v_D} \cos 2 \theta - \frac{c}{v_D} \cos 4 \theta \right)}{2  \sqrt{5 \left( \frac{c}{v_D} \right)^2 + 1 + 2 \frac{c}{v_D} \cos 3 \theta - 4 \left( \frac{c}{v_D} \right)^2 \cos 6 \theta}} \simeq - \frac{5 c}{16 \pi }, \\
    I^{(2,1)}_{yxy} &=& \int \frac{d \theta}{(2 \pi)^2} \frac{\mu}{v_D} \frac{1}{1 + \frac{c}{v_D} \cos 3\theta_{\bk}} \frac{1}{v_D \sqrt{5 \left( \frac{c}{v_D} \right)^2 + 1 + 2 \frac{c}{v_D} \cos 3 \theta - 4 \left( \frac{c}{v_D} \right)^2 \cos 6 \theta}} \left( \frac{\partial^2 \varepsilon(\bp)}{\partial p_{y} \partial p_{x}} \right) \left( \frac{\partial \varepsilon(\bp)}{\partial p_{y}} \right) = \nonumber \\
    &=& \int \frac{d \theta}{(2 \pi)^2} \frac{v_D (- \sin 2\theta) \left( 1 - 8 \frac{c}{v_D} \cos 3 \theta \right) \left( \sin \theta - 2 \frac{c}{v_D} \sin 2 \theta - \frac{c}{v_D} \sin 4 \theta \right)}{2  \sqrt{5 \left( \frac{c}{v_D} \right)^2 + 1 + 2 \frac{c}{v_D} \cos 3 \theta - 4 \left( \frac{c}{v_D} \right)^2 \cos 6 \theta}} \simeq - \frac{5 c}{16 \pi}, \\
    I^{(2,1)}_{xyy} &=& \int \frac{d \theta}{(2 \pi)^2} \frac{\mu}{v_D} \frac{1}{1 + \frac{c}{v_D} \cos 3\theta_{\bk}} \frac{1}{v_D \sqrt{5 \left( \frac{c}{v_D} \right)^2 + 1 + 2 \frac{c}{v_D} \cos 3 \theta - 4 \left( \frac{c}{v_D} \right)^2 \cos 6 \theta}} \left( \frac{\partial^2 \varepsilon(\bp)}{\partial p_{x} \partial p_{y}} \right) \left( \frac{\partial \varepsilon(\bp)}{\partial p_{y}} \right) = \nonumber \\
    &=& \int \frac{d \theta}{(2 \pi)^2} \frac{v_D (- \sin 2\theta) \left( 1 - 8 \frac{c}{v_D} \cos 3 \theta \right) \left( \sin \theta - 2 \frac{c}{v_D} \sin 2 \theta - \frac{c}{v_D} \sin 4 \theta \right)}{2  \sqrt{5 \left( \frac{c}{v_D} \right)^2 + 1 + 2 \frac{c}{v_D} \cos 3 \theta - 4 \left( \frac{c}{v_D} \right)^2 \cos 6 \theta}} \simeq - \frac{5 c}{16 \pi},
\end{eqnarray}
which is consistent with the 3-fold symmetry of the dispersion. In the same way one can explicitly check that the rest terms are zero.

Similarly, we can calculate the second integral
\begin{eqnarray}
    I^{(2,2)}_{xxx} &=& \int \frac{d \theta}{(2 \pi)^2} \frac{\mu}{v_D} \frac{1}{1 - \frac{c}{v_D} \cos 3\theta_{\bk}} \frac{1}{v_D \sqrt{5 \left( \frac{c}{v_D} \right)^2 + 1 - 2 \frac{c}{v_D} \cos 3 \theta - 4 \left( \frac{c}{v_D} \right)^2 \cos 6 \theta}} \left( \frac{\partial^2 \varepsilon(\bp)}{\partial p_{x} \partial p_{x}} \right) \left( \frac{\partial \varepsilon(-\bp)}{\partial p_{x}} \right) = \nonumber \\
    &=& \int \frac{d \theta}{(2 \pi)^2} \frac{v_D (1- \cos 2\theta) \left( 1 - 8 \frac{c}{v_D} \cos 3 \theta \right) \left( \cos \theta - 2 \frac{c}{v_D} \cos 2 \theta + \frac{c}{v_D} \cos 4 \theta \right)}{2  \sqrt{5 \left( \frac{c}{v_D} \right)^2 + 1 - 2 \frac{c}{v_D} \cos 3 \theta - 4 \left( \frac{c}{v_D} \right)^2 \cos 6 \theta}} \simeq \frac{11 c}{16 \pi }.
\end{eqnarray}
which also respects 3-fold rotational symmetry
\begin{eqnarray}
    I^{(2,2)}_{yyx} &=& \int \frac{d \theta}{(2 \pi)^2} \frac{\mu}{v_D} \frac{1}{1 - \frac{c}{v_D} \cos 3\theta_{\bk}} \frac{1}{v_D \sqrt{5 \left( \frac{c}{v_D} \right)^2 + 1 - 2 \frac{c}{v_D} \cos 3 \theta - 4 \left( \frac{c}{v_D} \right)^2 \cos 6 \theta}} \left( \frac{\partial^2 \varepsilon(\bp)}{\partial p_{y} \partial p_{y}} \right) \left( \frac{\partial \varepsilon(-\bp)}{\partial p_{x}} \right) = \nonumber \\
    &=& \int \frac{d \theta}{(2 \pi)^2} \frac{v_D (1+ \cos 2\theta) \left( 1 - 8 \frac{c}{v_D} \cos 3 \theta \right) \left( \cos \theta - 2 \frac{c}{v_D} \cos 2 \theta + \frac{c}{v_D} \cos 4 \theta \right)}{2  \sqrt{5 \left( \frac{c}{v_D} \right)^2 + 1 - 2 \frac{c}{v_D} \cos 3 \theta - 4 \left( \frac{c}{v_D} \right)^2 \cos 6 \theta}} \simeq - \frac{11 c}{16 \pi}, \\
    I^{(2,2)}_{yxy} &=& \int \frac{d \theta}{(2 \pi)^2} \frac{\mu}{v_D} \frac{1}{1 - \frac{c}{v_D} \cos 3\theta_{\bk}} \frac{1}{v_D \sqrt{5 \left( \frac{c}{v_D} \right)^2 + 1 - 2 \frac{c}{v_D} \cos 3 \theta - 4 \left( \frac{c}{v_D} \right)^2 \cos 6 \theta}} \left( \frac{\partial^2 \varepsilon(\bp)}{\partial p_{y} \partial p_{x}} \right) \left( \frac{\partial \varepsilon(-\bp)}{\partial p_{y}} \right) = \nonumber \\
    &=& \int \frac{d \theta}{(2 \pi)^2} \frac{v_D (- \sin 2\theta) \left( 1 - 8 \frac{c}{v_D} \cos 3 \theta \right) \left( \sin \theta + 2 \frac{c}{v_D} \sin 2 \theta + \frac{c}{v_D} \sin 4 \theta \right)}{2  \sqrt{5 \left( \frac{c}{v_D} \right)^2 + 1 - 2 \frac{c}{v_D} \cos 3 \theta - 4 \left( \frac{c}{v_D} \right)^2 \cos 6 \theta}} \simeq - \frac{11 c}{16 \pi}, \\
    I^{(2,2)}_{xyy} &=& \int \frac{d \theta}{(2 \pi)^2} \frac{\mu}{v_D} \frac{1}{1 - \frac{c}{v_D} \cos 3\theta_{\bk}} \frac{1}{v_D \sqrt{5 \left( \frac{c}{v_D} \right)^2 + 1 - 2 \frac{c}{v_D} \cos 3 \theta - 4 \left( \frac{c}{v_D} \right)^2 \cos 6 \theta}} \left( \frac{\partial^2 \varepsilon(\bp)}{\partial p_{x} \partial p_{y}} \right) \left( \frac{\partial \varepsilon(-\bp)}{\partial p_{y}} \right) = \nonumber \\
    &=& \int \frac{d \theta}{(2 \pi)^2} \frac{v_D (- \sin 2\theta) \left( 1 - 8 \frac{c}{v_D} \cos 3 \theta \right) \left( \sin \theta + 2 \frac{c}{v_D} \sin 2 \theta + \frac{c}{v_D} \sin 4 \theta \right)}{2  \sqrt{5 \left( \frac{c}{v_D} \right)^2 + 1 - 2 \frac{c}{v_D} \cos 3 \theta - 4 \left( \frac{c}{v_D} \right)^2 \cos 6 \theta}} \simeq - \frac{11 c}{16 \pi},
\end{eqnarray}
causing the other terms to vanish. 

Using the two calculated integrals we can write down an explicit expression for the 3-fold-symmetric component of the nonlinear conductivity tensor in our model:
\begin{equation}
    \tilde{\sigma}_{\text{3-fold}} = \frac{e^3 c}{16 \pi\left(  \frac{1}{\tau} - \alpha(0) \right) \left(  \frac{1}{\tau} + \alpha(0) \right)^2} \left[ \frac{5}{\tau} + 11 \alpha(0) \right].
    \label{nonlin_res}
\end{equation}

In the absence of weak (anti)localization effects $\alpha(0)=0$ and the ratio $\frac{\tilde{\sigma}}{\sigma^2}$ is independent of the scattering time:
\begin{equation}
    \frac{\tilde{\sigma}_{\text{3-fold}}}{\sigma^2} = \frac{1}{e}\frac{80 c \pi v_D^4}{\left( 9 c^2 + 4 v_D^2 \right)^2 \mu^2}.
\end{equation}
Note, that the sign of this quantity doesn't change with doping and depends only on the sign of $c$.

\subsection{Estimation of $\frac{\tilde{\sigma}}{\sigma^2}$ density dependence}
\label{app:estimation}

Using Eqs. \eqref{alpha_eq} and \eqref{eq_dos} we can estimate the density (chemical potential) dependence of $\frac{\tilde{\sigma}}{\sigma^2}$ to compare with experimental data from literature. Assuming independence of scattering time and $l_{\phi}$ on density we can get an expression for $\alpha(0)$ as a function of $\mu$:
\begin{equation}
    \alpha(0) = \pm \frac{\mathcal{C}}{\tau} \frac{4 v_D^4}{\pi \mu \left(4 v_D^2 - 3c^2 \right)} \frac{1}{D} \ln \frac{l_{\phi}}{l}.
\end{equation}
Substituting this expression into Eqs. \eqref{lin_res} and \eqref{nonlin_res} one can crudely estimate dependence of $\frac{\tilde{\sigma}}{\sigma^2}$ for $\mu>0$. Close to Dirac point this expression will fail and provide diverging linear conductivity in the weak antilocalization case because of artificial enhancement of $\alpha(0)$.

\section{Details of experimental data extraction and analysis}
\label{app:data_an}

\begin{figure*}[h!]
\includegraphics[width=0.99\linewidth]{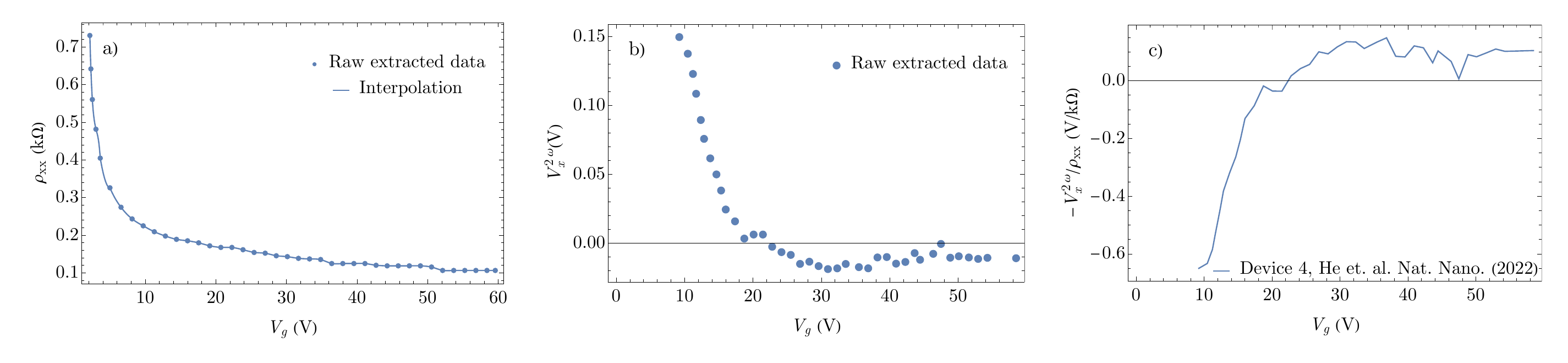}
\centering{}\caption{Data analysis details: a) Extracted $\rho_{xx}$ vs $V_g$ data from Supplementary Fig. 5d of Ref. \cite{He2022} (scatter points) and interpolation of the data (continuous line); b) Extracted $\tilde{\sigma}_{xxx}$ vs $V_g$ data from Supplementary Fig. 5e of Ref. \cite{He2022}; c) Calculated  $- \frac{V_{x}^{2 \omega}}{\rho_{xx} }$ using raw data from panel b) and values of interpolation function from panel a). Panel c) is presented in the main text.
} 
\label{fig_SI_data}
\end{figure*}

To compare with experimental data we extract values of nonlinear potential $V_x^{2 \omega}$ and of linear resistivity $\rho_{xx}$ as a function of gate voltage $V_{g}$ (a proxy to electron density) from the Supplementary Materials of Ref. \cite{He2022}. Specifically, we use data for Device 4 from Supplementary Fig. 5d,e for positive $V_g$, which we attribute to electron doping of the system. 
\end{widetext}

To plot $\tilde{\sigma}_{xxx}/\sigma_{xx}^2$ as a function of $V_g$ we need to express this quantity via $V_x^{2 \omega}$ and $\rho_{xx}$. To do that, we follow the procedure described in the Methods section of Ref. \cite{He2022}. There, the authors used second harmonic voltage and linear voltage drops to relate current densities at second harmonic frequency:
\begin{eqnarray}
    j_{x}^{2 \omega} &=& \tilde{\sigma}_{xxx} \left(E_0 \right)^2, \nonumber \\
    j_{x}^{2 \omega} &=& \sigma_{xx} E_x^{2 \omega}, \nonumber \\
    V_0 &=& -E_0 L, \nonumber \\
    V_{x}^{2 \omega} &=& -E_{x}^{2 \omega} L, \nonumber
\end{eqnarray}
where $E_0$ and $V_0$ are the electric field and voltage at frequency $\omega$, $j_{x}^{2 \omega}$ and $E_x^{2 \omega}$ are current density and electric field at frequency $2\omega$ in the $x-$direction, $L$ is the linear dimension of the sample. The measurement is performed for a fixed injected current magnitude and at low $\omega$ so that the current is essentially in the DC limit. From the equations above one can relate $\tilde{\sigma}_{xxx}$ and $\sigma_{xx}$:
\begin{equation}
    \tilde{\sigma}_{xxx} = - \frac{\sigma_{xx} V_{x}^{2 \omega} L}{ V_0^2},
\end{equation}
which is the same expression as in Ref. \cite{He2022}. Assuming perfect 3-fold symmetry of the system we use $\rho_{xx} = \frac{1}{\sigma_{xx}}$ and 
$$
j_{x}^{\omega} = \sigma_{xx} E_0
$$
to express 
\begin{equation}
    \frac{\tilde{\sigma}_{xxx}}{\sigma_{xx}^2} = - \frac{\sigma_{xx} V_{x}^{2 \omega}}{\left( j_{x}^{\omega} \right)^2 L} \propto - \frac{V_{x}^{2 \omega}}{\rho_{xx} }.
\end{equation}
Since we're after the mere shape of the curve and not the exact magnitude, we plot 
$- \frac{V_{x}^{2 \omega}}{\rho_{xx} }$ as function of $V_g$. To make sure that the values of $\tilde{\sigma}_{xxx}$ and $\sigma_{xx}$ in the evaluation of $\frac{\tilde{\sigma}_{xxx}}{\sigma_{xx}^2}$ correspond to the same $V_g$, we interpolate $\sigma_{xx} (V_g)$ and calculate $\frac{\tilde{\sigma}_{xxx}}{\sigma_{xx}^2}$ using this interpolation function to evaluate $\sigma_{xx}$ for the same values of $V_g$ as in the raw extracted dataset for $\tilde{\sigma}_{xxx}$, see Fig. \ref{fig_SI_data}.

\subsection{Additional data from Ref. \cite{Liu2024experiment}}

Here we show an additional set of data extracted from Fig. 4a,b of Ref. \cite{Liu2024experiment} for $T=40$K, which is above the N\'eel temperature $T_N \sim 23$K. One can see a trend, similar to the one discussed in the main text. 

\begin{figure}[h!]
\includegraphics[width=0.99\linewidth]{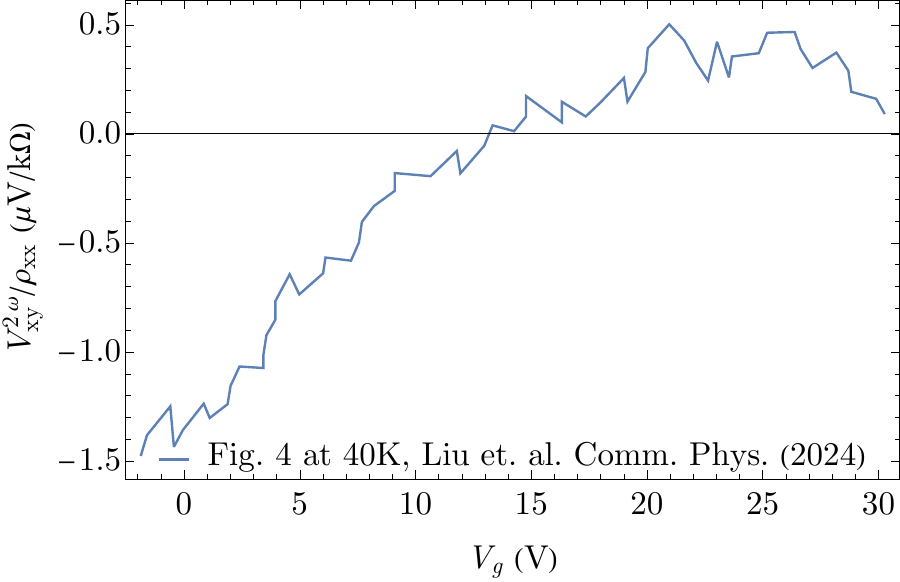}
\centering{}\caption{
{Experimental data extracted from Ref. \cite{Liu2024experiment} following the analysis procedure discussed in the Appendix \ref{app:data_an}. }
} 
\label{fig_additional_data}
\end{figure}

\bibliography{biblio}

\end{document}